\documentclass{PoS}
\usepackage{amsfonts}
\usepackage{mathpazo}
\usepackage{longtable}
\usepackage{colordvi}
\usepackage{graphicx,epsfig}
\usepackage{amsmath,amssymb}

\title{Using X-ray catalogues to find counterparts to unassociated high-energy \emph{Fermi}/LAT
sources}

\ShortTitle{Using X-ray catalogues to find counterparts to unassociated high-energy Fermi/LAT
sources}

\author{\speaker{R. Landi}\\
        INAF -- IASF Bologna\\
        E-mail: \email{landi@iasfbo.inaf.it}}

\author{L. Bassani\\
        INAF -- IASF Bologna\\
        E-mail: bassani@iasfbo.inaf.it}

\author{J. B. Stephen\\
        INAF -- IASF Bologna\\
        E-mail: stephen@iasfbo.inaf.it}

\author{N. Masetti\\
        INAF -- IASF Bologna\\
        E-mail: masetti@iasfbo.inaf.it}
        
\author{A. Malizia\\
        INAF -- IASF Bologna\\
        E-mail: malizia@iasfbo.inaf.it}

\author{P. Ubertini\\
        INAF -- IAPS Rome\\
        E-mail: pietro.ubertini@iaps.inaf.it}

\abstract{
The first \emph{Fermi} Large Area Telescope (LAT) catalogue of sources (1FHL) emitting at high energies 
(above 10 GeV) reports the 
details of 514 objects detected in the first three years of the \emph{Fermi} mission. Of these, 71 were 
reported as unidentified in the 1FHL catalogue, although six are likely to be associated with a supernova 
remnant (SNR), a Pulsar Wind Nebula (PWN) or a combination of both, thereby leaving a list of 65 still 
unassociated objects. Herein, we report a preliminary analysis on this sample of objects concentrating 
on nine 1FHL sources, which were found to have a clear optical extragalactic classification. They are 
all blazar, eight BL Lac and one flat spectrum radio quasar, typically at redshift greater than 0.1.
}

\FullConference{Swift: 10 Years of Discovery, \\
		2--5 December 2014\\
		La Sapienza University, Rome, Italy}

\begin{document}

\section{Introduction}
We report on a preliminary work that is part of a larger programme aimed at identifying  
unassociated sources in the first \emph{Fermi}/LAT high-energy catalogue [1FHL, 1], which contains
514 objects detected above 10 GeV. The majority of these sources are identified with known objects
(449 or 87\% of the sample): approximately 75\% with AGNs (mostly blazars), while Galactic sources
(pulsars, PWNs, SNRs, high-mass binaries, and star-forming regions) collectively represent 10\%
of the sample. The fraction of unassociated sources is less than 14\% corresponding to 71   
objects, of which six are likely to be associated with a SNR, a PWN or a combination
of both, thereby leaving a list of 65 still unidentified objects. The third \emph{Fermi}/LAT
catalogue [3FGL, 2], which has recently been published, contains most of these
unassociated 1FHL sources except for 13 objects that are missing. The main motivation behind the
1FHL catalogue was to find the hardest gamma-ray sources in the sky and to get a sample of
objects which are good candidate for detection at TeV energies.

As a first step, we have cross-correlated the sample of 65 objects with both the \emph{ROSAT} Bright 
(RASSBSC, [3]) and the \emph{XMM-Newton} Slew Survey [4] catalogues, following the 
prescription of [5] and finding the likely counterpart to 19 1FHL sources. Secondly, 
we have extended our analysis using data collected with the X-ray telescope (XRT) on-board \emph{Swift} 
[6]; this was done by cross-correlating the list of unassociated 1FHL sources with all 
the XRT archival data up to the end of 2014 and found to be within around 10 arcmin from the 
\emph{Fermi} best-fit position. This analysis has led us to investigate a further set of sources, 
increasing the 
sample for which a likely association is found to around 30, i.e. half of the original set of objects. 
The remaining 1FHL sources have also been investigated on an individual basis. The nature of each likely 
counterpart has been studied by means of a multi-waveband approach using information in the radio, 
infrared, and optical wavebands. 
In particular, we use the WISE colours as discussed by [7] to 
test the possible blazar nature of each source: these authors found that in the $W2-W3$ versus $W1-W2$ 
colour-colour plot, the positions of gamma-ray emitting blazars are all within a well-defined region 
known as the ``Blazar strip''.

Herein, we report some results from this on going programme concentrating on nine 1FHL objects, which were 
found to have an optical classification. All sources have a counterpart in the third \emph{Fermi}/LAT
catalogue and the same association we found in this work, although two display multiple X-ray counterparts 
and other two an X-ray detection outside the 1FHL 95\% positional uncertainty.

\begin{table*}[t]
\begin{center}
\scriptsize
\caption{Unidentified \emph{Fermi} 1FHL sources with an RASSBSC/XMMSlew/XRT counterpart.}
\begin{tabular}{lccccc}
\hline
\hline
\multicolumn{1}{c}{\emph{Fermi} Name} & \multicolumn{2}{c}{X-ray counterpart} & X-ray error$^\dagger$ & 
Catalogue & Optical class ($z$) \\
    & R.A.(J2000) & Dec.(J2000) & (arcsec)  &   &   \\
\hline
1FHL J0110.0$-$4023    & 01 09 56.5 & $-$40 20 47.0  & 7.0  & RASSBSC &  BL Lac (0.313)    \\
1FHL J0118.5$-$1502    & 01 19 05.4 & $-$14 59 06.0  & 14.0 & RASSBSC &  BL Lac (0.1147)   \\
1FHL J0601.0$+$3838$^\ddagger$  & 06 01 02.7 & $+$38 38 27.2 & 5.2 & XRT  & BL Lac   \\         
1FHL J0828.9$+$0902    & 08 29 30.1  & $+$08 58 20.5 & 4.2 & XRT    &  FSRQ   (0.866)    \\
1FHL J0841.2$-$3556$^\ddagger$  & 08 41 21.6  & $-$35 55 50.8 & 6.0 & XRT & BL Lac ($\ge$ 0.15) \\
1FHL J1353.0$-$6642$^\ddagger$  & 13 53 41.1  & $-$66 40 02.0 & 8.0 & RASSBSC & BL Lac ($\ge$ 0.15)\\
                                & 13 53 40.6  & $-$66 39 58.0 & 3.0 & XMMSlew &   --    \\
1FHL J1406.4$+$1646             & 14 06 59.2  & $+$16 42 06.0 & 3.7 & XRT & BL Lac ($\ge$ 0.623) \\
1FHL J1440.6$-$3847             & 14 40 37.4  & $-$38 46 58.5 & 7.0 & RASSBSC & BL Lac   \\
                                & 14 40 38.1  & $-$38 46 53.8 & 3.0 & XMMSlew &  --   \\ 
1FHL J2004.7$+$7003             & 20 05 04.5  & $+$70 04 40.6 & 4.0 & XMMSlew & BL Lac              \\
\hline
\hline
\end{tabular}  
\begin{list}{}{}
\item $^\dagger$ \emph{ROSAT}, \emph{XMM-Newton} Slew errors are 1$\sigma$ radius, while \emph{Swift}/XRT 
errors are 1.6$\sigma$ radius; $^\ddagger$ Object at low Galactic latitude, i.e. within $\pm$10 degrees 
of the Galactic plane.
\end{list}
\end{center}
\end{table*}

\begin{figure*}
\centering
\includegraphics[width=0.35\linewidth]{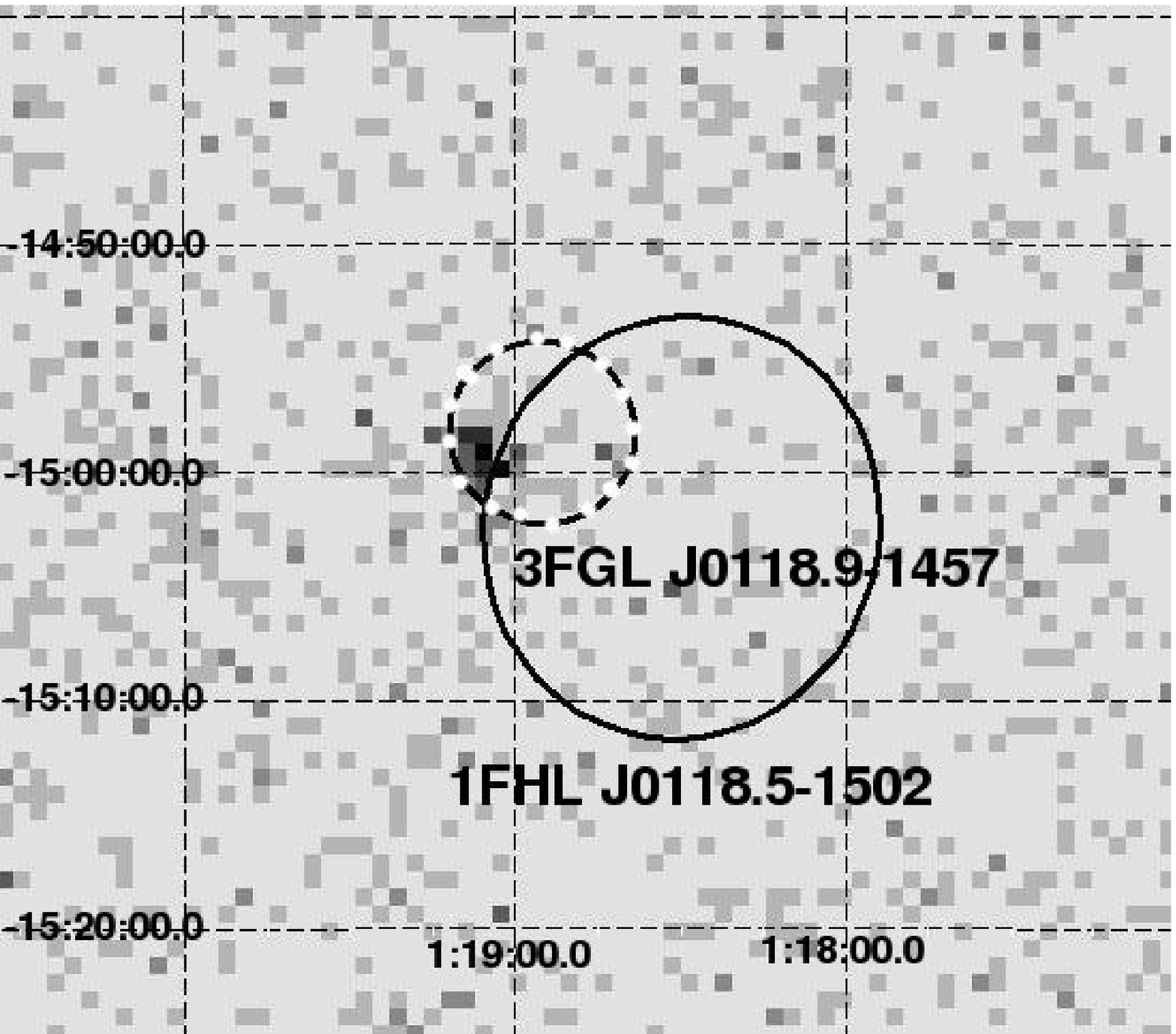}
\includegraphics[width=0.4\linewidth]{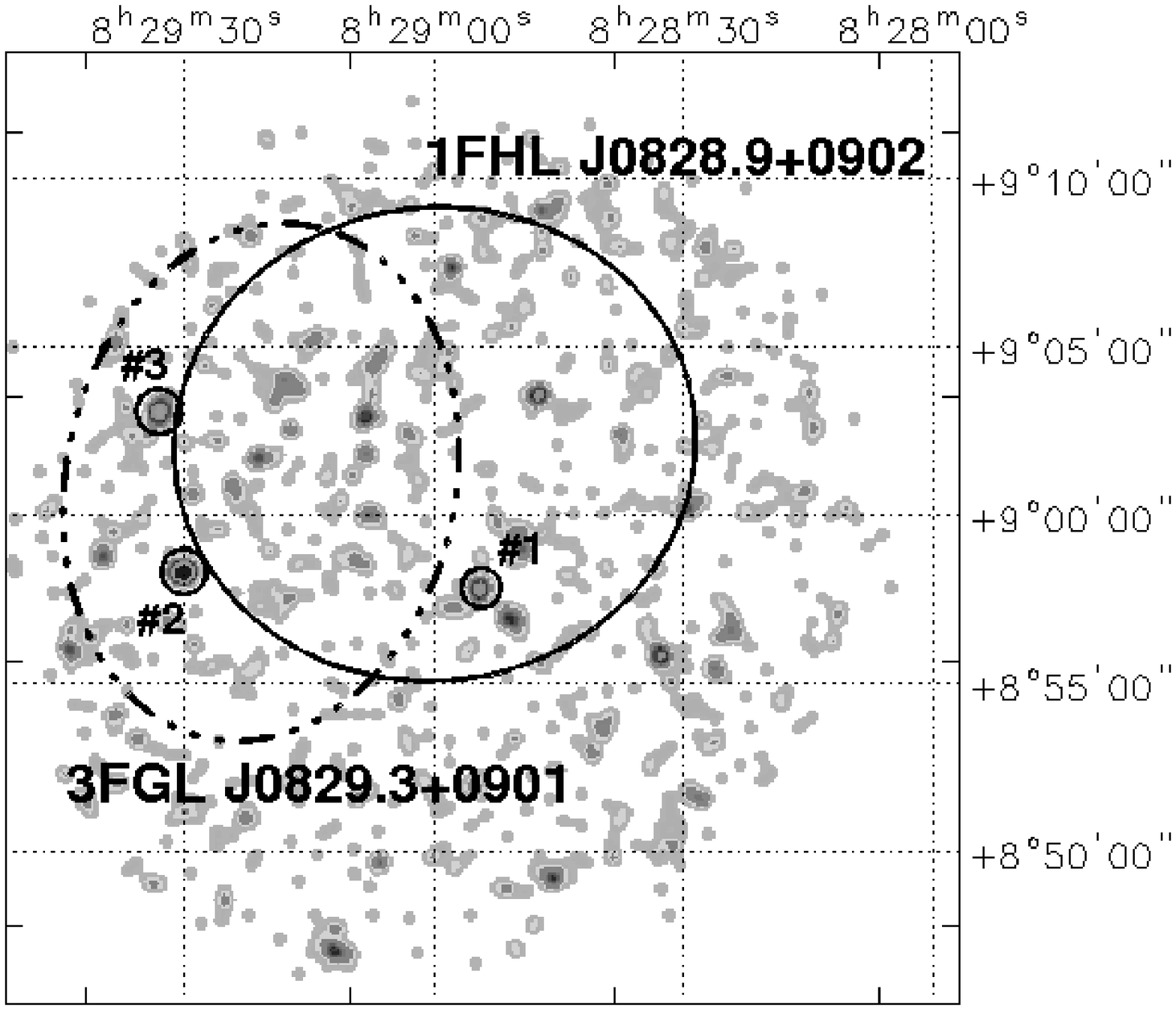}
\includegraphics[width=0.4\linewidth]{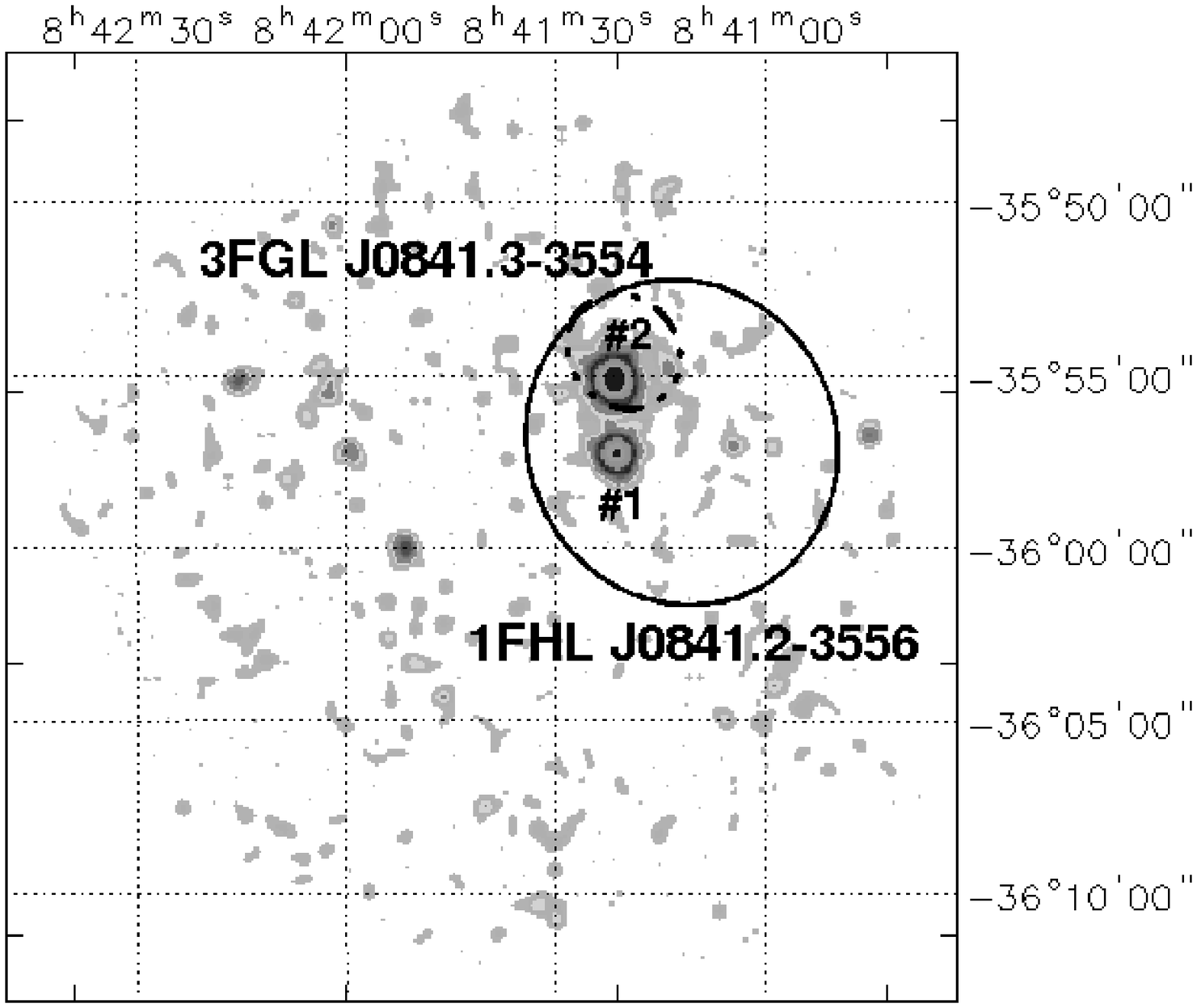}
\includegraphics[width=0.4\linewidth]{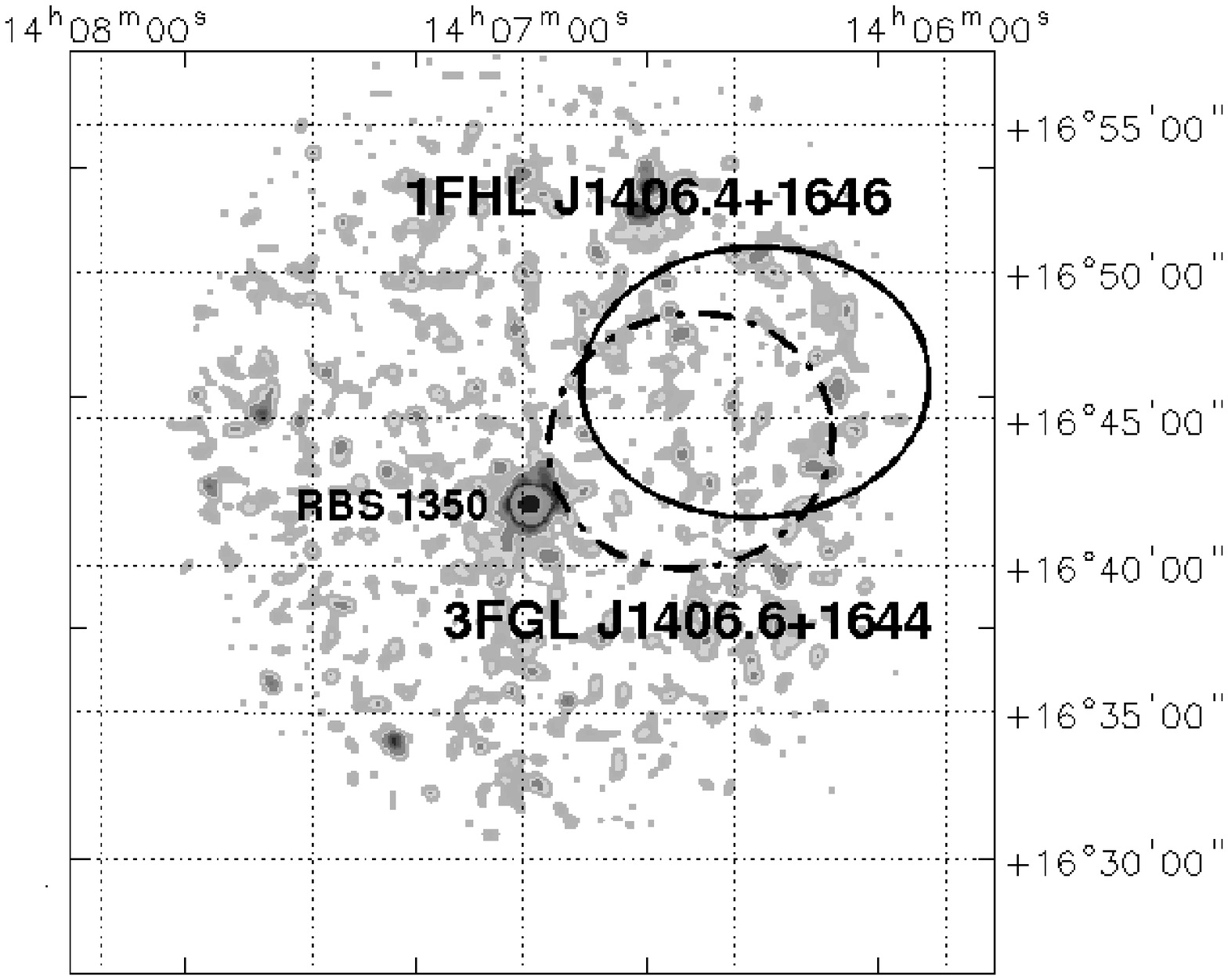}
\caption {X-ray images of 1FHL J0118.5--1502 (\emph{ROSAT}, upper left panel),
1FHL J0841.2--3556 (XRT, upper right panel), 1FHL J0828.9+0902 (XRT, bottom left panel), and 
1FHL 1406.4+1646 (XRT, bottom right panel). The black ellipse and the black-dotted ellipse depict the 
positional uncertainty of the 1FHL and 3FGL sources, respectively. See details in the text.}
\label{f1}
\end{figure*}

\subsection{Objects classified optically}

Nine of the various sources analysed in the project were found to have a likely association with 
known blazars: they are listed in Table 1, where we 
report the \emph{Fermi} name, the coordinates of the associated soft X-ray counterpart we found (from 
either \emph{ROSAT} Bright, \emph{XMM-Newton} Slew or \emph{Swift}/XRT observations), the X-ray 
error radius, the source optical class, and the redshift when available. In the following, we describe 
briefly each individual source.

The soft X-ray counterpart to 1FHL J0110.0--4023 is associated with RBS0158 (also ATESP
J010956--402051) a radio source that shows 20 and 36 cm flux densities of 57 and 36 mJy, respectively 
[8]. The source, which was optically classified as a BL Lac object by [9], has a
redshift of 0.313. It is also listed in the WISE catalogue [10] with colours 
$W2-W3=1.84$ and $W1-W2=0.59$, which are well inside the blazar strip.

The small XRT error circle of 1FHL J0601.0$+$3838 allows the identification of the X-ray source
with the bright radio object B20557+38, which displays 20 and 92 cm flux densities of 704 and 1882
mJy, respectively. This object is reported in various radio archives and has a radio spectrum with
index of $\sim$0.7 (see NASA/IPAC Extragalactic Database, NED). The source has WISE colours
$W2-W3=2.47$ and $W1-W2=0.97$, typical of gamma-ray emitting blazars. It was optically classified as a 
BL Lac by [11].

For 1FHL J1353.0--6642, the restricted X-ray position provides a secure identification with VASC
J1353--66. This object, which is listed in the \emph{XMM-Newton} Slew Survey, has an X-ray 0.2--12 
keV 
flux of $3.9\times 10^{-12}$ erg cm$^{-2}$ s$^{-1}$.
It is detected in radio at various frequencies, including the 36 cm one (flux density of 70.7 mJy, see 
[12]), and shows a flat radio spectrum (see [13]), while it is not detected by
WISE. The source was optically classified as a BL Lac by [13], while [14]
were able to put a lower limit of 0.15 to the source redshift.

The X-ray counterpart to 1FHL J1440.6--3847 is unambiguously identified with the galaxy 6dF
J1440378--384655, which is detected at 20 and 36 cm with flux densities of 22.8 and 23.2 mJy,
respectively; the 0.2--12 keV flux is $7.9\times 10^{-12}$ erg cm$^{-2}$ s$^{-1}$. The WISE
colours ($W2-W3=1.38$ and $W1-W2= 0.62$) locate the source outside the blazar strip.
6dF J1440378--384655 is classified as a BL Lac in 
NED (see also [15]), but on the basis of a poor quality optical spectrum.

The soft X-ray counterpart to 1FHL J2004.7$+$7003 is radio detected at 20 cm with a flux density of 6.5
mJy and listed in the WISE catalogue with colours $W2-W3=2.21$ and $W1-W2=0.77$, i.e. fully 
compatible with the blazar strip. It is variable in
both WISE ($W1$ and $W2$ wavebands) and \emph{XMM-Newton} Slew catalogues: the X-ray 0.2--12 keV 
flux ranges from
2.8 to $8.2 \times10^{-12}$ erg cm$^{-2}$ s$^{-1}$. This source was studied and discussed by
various authors: all suggested that it is probably a BL Lac (see [16];[17];[18]), as confirmed in the 
3FGL catalogue.

Four cases deserve a more in-depth analysis because they have multiple X-ray counterparts or have an 
association located outside the 1FHL positional uncertainty.

The only X-ray source we found in the case of 1FHL J0118.5--1502 is a bright \emph{ROSAT} source, which is 
located just outside the border of the \emph{Fermi} error ellipse (left upper panel of Figure 1).
Despite this, the source is within the 
positional uncertainty quoted for the 3FGL counterpart. The \emph{ROSAT} source, which has the greatest 
error radius reported in Table 1, has a radio association in the NVSS (NVSS J011904--145858) with a 20 cm 
flux density of 5 mJy and is WISE-detected with colours $W1-W2=0.521$ and $W2-W3=1.713$. The 
source was optically studied by [15] and found to display a spectrum with only absorption 
lines: it was therefore classified as a BL Lac at redshift 0.1147.

In the case of 1FHL J0828.9+0902, various soft X-ray sources are found inside or at the border 
of the 1FHL error ellipse (right upper panel of Figure 1).
Source \#1 is not detected in radio and has WISE colours that are not 
compatible with a blazar classification. Despite this, it is listed in NED as a QSO candidate (SDSS 
J082854.54+085751.2) at $z=0.855$ (see [19]). Similarly, source \#3 is not detected at 
radio frequencies, and it does not show WISE colours compatible with those typically 
displayed by blazars. 
The remaining object (source \#2) coincides with the radio source NVSS J082930+085821 (also TXS 
0826+091), which displays a 20 cm flux density of 333.9 mJy. This X-ray detection is identified with a 
QSO at 
$z=0.866$ in NED. It was also classified as a flat spectrum radio object by [20], but its
WISE colours ($W2-W3=3.08$ and $W1-W2=0.66$) are outside the blazar strip. Source \#2 is also the 
association reported in the 3FGL catalogue and it looks like the most promising one at the moment.

The error box of 1FHL J0841.4--3558 contains two X-ray sources, as evident in the left lower panel of 
Figure 1: one is a bright \emph{ROSAT}/XRT object 
(R.A.(J2000) = $08^\text{h}41^\text{m}21^{\prime}.40$ and Dec.(J2000) = 
$-35^\circ57^{\prime}04^{\prime \prime}.50$, 9 arcsec error radius) associated with the star 2MASS 
J08412132--3557154 (also HIP 42640) of spectral type F2V, which is unlikely to emit gamma-rays. The other, 
reported in Table 1, is located only 1.3 arcmin north of the star; it is listed as a radio source in 
various catalogues and has WISE colours $W2-W3=2.328$ and $W1-W2=0.863$, which locate the source 
in the locus of gamma-ray blazars. The optical spectrum 
obtained recently by [14] is featureless and the source was classified as a BL Lac 
at $z>0.15$. Note that this source is still listed as an unclassified blazar in the third 
\emph{Fermi}/LAT catalogue.

In the case of 1FHL J1406.4+164, the only X-ray detection is just 3 arcmin outside the \emph{Fermi} 1FHL 
error ellipse, but it is located at the border of the positional uncertainty of 3FGL J1406.6+1644 
(right lower panel of Figure 1).
This X-ray object is associated with RBS 1350, which is classified as a BL Lac object and 
suggested to be an extreme high-energy peaked blazar or a TeV candidate
(see [21]; [22]). The source
redshift has a lower limit of 0.623 and a photometric value
of 1.985. In radio, the source has a 20 cm flux density around 78 mJy,
while its WISE colours ($W2-W3=2.362$ and $W1-W2=0.574$) are compatible with the blazar 
strip. Given the overall properties RBS 1350 and the overlap in positional uncertainties between the 
1FHL and the 3FGL sources, we regard the association proposed here as likely, although not certain.

\begin{figure*}
\centering
\includegraphics[width=6.0cm,height=8.5cm,angle=-90]{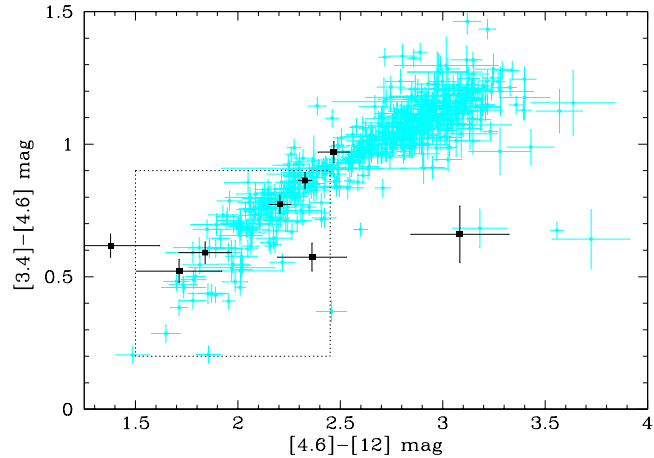}
\caption {The [4.6]-[12]/[3.4]-[4.6] MIR colour-colour plot reporting the positions of
gamma-ray emitting blazars (in cyan) associated with WISE
sources forming the blazar strip (see [7] for more details), together with the
BL Lac objects (filled squares) identified in this paper.}
\label{f2}
\end{figure*}

\section{Discussion and conclusions}

The main result of this work is that we have associated nine unidentified 1FHL sources with 
blazars, eight of the BL Lac type and one of the Flat Spectrum Radio Quasar type. Another interesting 
result 
is that all these sources are at redshift higher than 0.1 and hence allow probing the BL Lac population at 
a further distance than usual. The third evidence, coming from this work, is that all our BL 
Lacs are good candidates to be TeV emitting objects. As discussed by [14], and in the references 
therein, the TeV emitting BL Lacs populate a well-defined region of the WISE colour-colour 
diagram, i.e a square 
located in the lower part of the blazar strip. Therefore, objects with colours compatible with the 
TeV square are good candidates to emit at TeV energies. The 
WISE colour-colour diagram for the objects listed in Table 1, for which we have 
WISE colours, is plotted in Figure 2: as expected, all BL Lacs lie within or nearby 
the limits of the locus populated by TeV-emitting BL Lacs and therefore they are good candidate for very
high-energy observations; the only exception is the FSRQ (1FHL J0828.9+0902), which is even located 
outside the blazar strip.
Further results stemming from the analysis described above are being prepared, and an optical follow-up 
program of the associations we found is well underway with time already assigned at various telescopes. 
Overall, the results obtained so far validate the goodness of our analysis method, which can be applied to 
the much larger set of still unidentified sources in the 3FGL catalogue or in further high-energy 
\emph{Fermi} catalogues.

\end{document}